\documentstyle[floats,twocolumn,aps,prl,epsf,rotate] {revtex}

\begin{document}
\draft
\tighten

\title{Hydrogen dissociation on metal surfaces -- a model system for
       reactions on surfaces} 

\author{Axel Gro{\ss}}

\address{Fritz-Haber-Institut der Max-Planck-Gesellschaft, Faradayweg 4-6, 
         D-14195 Berlin-Dahlem, Germany
         }

\maketitle

\begin{abstract}

Reactions on surfaces play an important role in many technological 
applications. Since these processes are often rather complex,
one tries to understand single steps of these complicated reactions 
by investigating simpler system. In particular the hydrogen dissociation
on surfaces serves as such a model system.
There has been much progress in recent years in the theoretical
description of reactions on surfaces by high-dimensional dynamics
simulations on potential energy surfaces which are derived
from {\it ab initio} total energy calculations. In this brief review
I will focus on  the hydrogen dissociation on the clean
and sulfur-covered Pd(100) surface. These calculations established
the importance of dynamical concepts like the steering effect.
The electronic structure calculations allow furthermore the determination
of the factors that determine the reactivity of a particular surface.
This will be demonstrated for the poisoning of hydrogen dissociation
by sulfur adsorption on the Pd(100) surface.
In addition, quantum effects in the dynamics can be assessed
by comparing the results of classical with quantum 
dynamical calculations on the same potential energy surface.

\end{abstract}


\section{Introduction}

Reactions on surfaces play an important role in many technological 
applications like the heterogenous catalysis, growth of semiconductor 
devices, corrosion and lubrication of mechanical parts, or hydrogen
storage in metals. In spite of this significance the understanding
of the microscopic details of these reactions is still rather incomplete.
Of particular importance are processes in which chemical bonds of
molecules are breaking due to the presence of a substrate because
there processes represent the first elementary step in, e.g.,
heterogeneous catalysis or corrosion. Often this is the rate-limiting
step, for example in the ammonia synthesis. Hydrogen dissociation
on metal surfaces has become {\em the} model system for the
bond-breaking process on surfaces in the last years
because it can be studied in detail experimentally as well as
theoretically. In particular in the theoretical description
there has been much progress recently due to the improvement of 
computer power and the development of efficient algorithms.
It has become possible to map out detailed potential energy surfaces
of the dissociation of hydrogen on metal surfaces 
by density functional theory calculations  
\cite{Ham94,Whi94,Wil95,Wil96PRB,Whi96PRB,Wie96,Eich96,Dong96}.
The availability of high-dimensional
reliable potential energy surfaces has challenged the dynamics community
to improve their methods in order to perform high-dimensional dynamical 
studies on these potentials. Now quantum studies of the dissociation of 
hydrogen on surfaces are possible in which all six degrees of 
freedom of the molecule are treated dynamically
\cite{Gro95PRL,Gro98PRB,Kro97PRL,Kro97JCP,Dai97}.
In this brief review I will illustrate this progress by focusing
on the hydrogen dissociation on the clean and sulfur-covered
palladium surface. 

Hydrogen is the simplest molecule which
makes it accessible to a relatively complete theoretical treatment.
At the same time hydrogen is also well-suited for performing
experiments which allows a fruitful interaction between theory
and experiment.  I will show that general concepts relating to
the reactivity of surfaces as well as to dynamical reaction
mechanisms can be deduced from the detailed comparison
of theoretical and experimental results of the hydrogen dissociation
of metal surfaces. These concepts are applicable to any reaction
system making hydrogen the ideal candidate
for studying reactions on surfaces.

\section{General concepts in the adsorption dynamics at surfaces}

The sticking or adsorption probability is defined as the fraction
of atoms or molecules impinging on a surface that are not
scattered back, i.e. that remain on the surface. It should be noted
here that there is no unambiguous definition of the sticking probability
because for surfaces with non-zero temperature every adsorbed particle
will sooner or later desorb again. Hence the sticking probability
depends on the time-scale of the required residence time on the surface.

Atomic adsorption is often very efficient. 
Hydrogen atoms, e.g, stick at metal surfaces \cite{Eilm96} and 
semiconductor surfaces \cite{Schu83} with a probability of order unity. 
However, dissociative adsorption probabilities can differ by many orders 
of magnitude. Whereas the sticking probability of hydrogen molecules on 
many transition metal surfaces is about 0.5 \cite{Eilm96,Ren89},
at room temperature the dissociation probability of H$_2$/Si
is only 10$^{-8}$ \cite{Bra96PRB}, and for N$_2$/Ru it is even as low as
10$^{-13}$ \cite{Hin97}. The investigation of processes that occur within
such a wide range of probabilities represents of course a great challenge
for the theory as well as for the experiment.

\begin{figure}[t]
\unitlength1cm
\begin{center}
   \begin{picture}(10,6.5)
\put(-1.,0.){ \rotate[l]{\epsfysize=10.cm  
          \epsffile{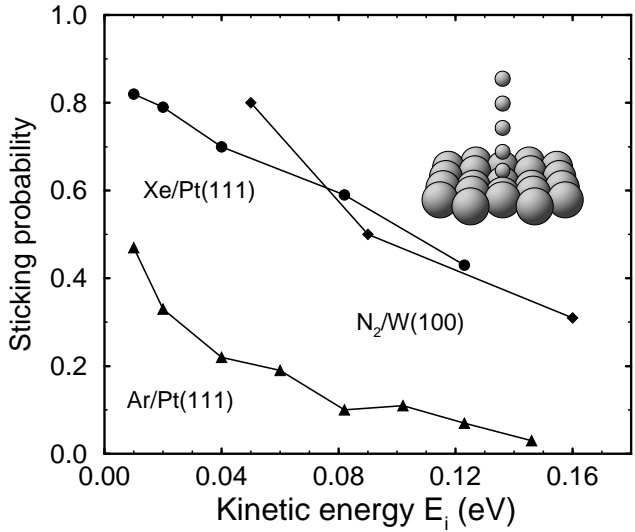}} }
   \end{picture}

\end{center}
   \caption{Atomic adsorption probability for Xe and Ar on Pt(111) and
            molecular adsorption probability of N$_2$/W(100). These
            examples are taken from the textbook by Zangwill 
            \protect\cite{Zan88}. The inset illustrates the adsorption
            process.
            }            

\label{mol_ads}
\end{figure}

There is one fundamental difference between atomic and molecular adsorption 
on the one side and dissociative adsorption on the other side that is very
important for the theoretical description of these processes. Here with 
molecular adsorption a sticking process is meant in which the molecule 
stays intact on the surface. This fundamental difference will be illustrated
in the following. In atomic and molecular adsorption it is crucial that
the impinging particles transfer their kinetic energy to the surface,
otherwise they would be scattered back into the gas phase.
If $P_E (\epsilon)$ is the probability that an incoming particle with
kinetic energy $E$ will transfer the energy $\epsilon$ to the surface,
then the atomic or molecular sticking probability can be expressed as
\begin{equation}
S(E) \ = \ \int_{E}^{\infty} \ P_E(\epsilon) \ d\epsilon,
\end{equation}
i.e., it corresponds to the fraction of particles that transfer more
energy to the surface than their initial kinetic energy. This excess
energy has to be transferred to substrate excitation, i.e., either
phonons or electron-hole pairs. Hence any theoretical description
of atomic or molecular adsorption has to consider dissipation to the continous
excitation spectrum of the substrate. In Fig.~\ref{mol_ads}
sticking probabilities for atomic and molecular adsorption as a function
of the initial kinetic energy are shown that correspond indeed to textbook 
examples \cite{Zan88}. These curves show a typical behavior, namely the
decrease of the sticking probability with increasing kinetic energy.
This is due to the fact that the energy transfer to the surface becomes
less efficient at higher kinetic energies. Of course, the higher the kinetic 
energy is, the  more energy is transfered to the surface. But the fraction
of particles that loose more energy than their initial kinetic energy
becomes smaller at higher kinetic energy. There is still more
interesting physics in these sticking probabilities. For example, at low 
kinetic energies classically the sticking probability should become
unity if there is no barrier before the adsorption well. Every
impinging particle transfers energy to the substrate so that
in the limit of zero initial kinetic energy all particles will stick.
Quantum mechanically, however, there is a non-zero probability for
elastic scattering at the surface so that the sticking probabilities
become less than unity in the zero-energy limit \cite{Sch88}. In 
Fig.~\ref{mol_ads} these quantum effects at low energies are evident in 
the sticking probability of the light noble gas argon on Pt(111) compared
to the sticking probability of the heavier noble gas xenon on the same
surface.

\begin{figure}[t]
\unitlength1cm
\begin{center}
   \begin{picture}(10,6.5)
\put(-1.,0.){ \rotate[l]{\epsfysize=10.cm  
          \epsffile{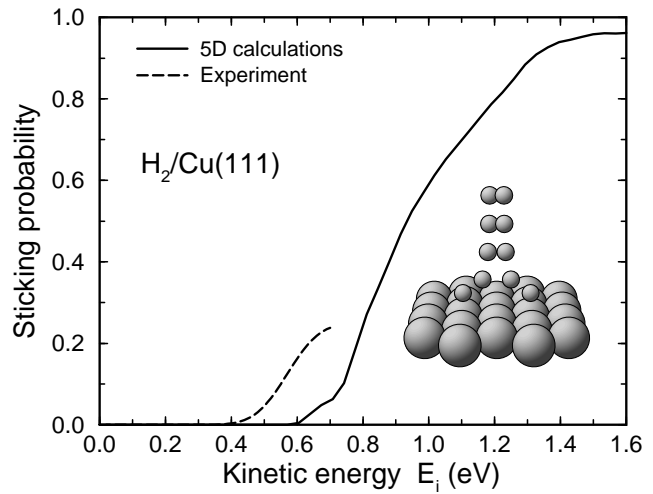}} }
   \end{picture}

\end{center}
   \caption{Dissociative adsorption probability versus kinetic energy 
     of H$_2$/Cu(111) for molecules initially in the vibrational ground state. 
  Solid line: results of five-dimensional calculations in which the molecular
            axis was kept parallel to the surface 
           (from ref.~\protect\cite{Gro94PRL}).
     Dashed line: Experimental curve (from ref.~\protect\cite{Ret95}).
     The inset illustrates the dissociation process. }

\label{diss_ads}
\end{figure}

Now in the case of dissociative adsorption there is another channel
for energy transfer, which is the conversion of the kinetic and internal 
energy of the molecule into translational energy of the atomic fragments 
on the surface relative to each other. This represents the fundamental
difference to atomic or molecular adsorption. It is true that eventually
the atomic fragments will also dissipate their kinetic energy and come
to rest at the surface. However, especially in the case of light molecules
like hydrogen dissociating on metal surfaces the energy transfer to the
substrate is very small due to the large mass mismatch. Whether a molecule
sticks on the surface or not is almost entirely determined by the 
bond-breaking process for which the energy transfer to the substrate
can be neglected. This makes it possible to describe the dissociative
adsorption process within low-dimensional potential energy surfaces
neglecting the surface degrees of freedom
if furthermore no substantial surface rearrangement upon adsorption occurs,
as it is usually case in the dissociative adsorption on close-packed
metal surfaces. Fig.~\ref{diss_ads} shows the dissociative adsorption 
probability of a system which also corresponds to a textbook example,
namely the dissociative adsorption of H$_2$ on Cu(111). In this system
the dissociation is hindered by a noticeable barrier so that the
dependence of the sticking probability on the kinetic energy exhibits
a behavior typical for activated systems \cite{Gro94PRL,Ret95}.

\section{Dissociative adsorption at a transition metal surface}

\begin{figure}[tb]
\unitlength1cm
\begin{center}
   \begin{picture}(10,6.5)
\centerline{   \rotate[r]{\epsfysize=8.cm  
          \epsffile{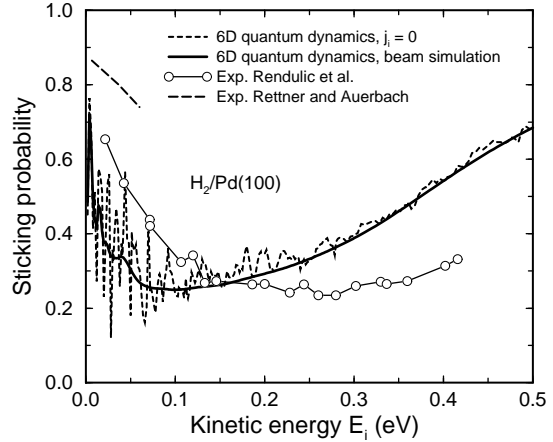}} }

   \end{picture}
\end{center}
   \caption{Sticking probability versus kinetic energy for
a hydrogen beam under normal incidence on a Pd(100) surface.
Theory: six-dimensional results for H$_2$ molecules initially in the 
rotational and vibrational ground state (dashed line)
and with an initial rotational and energy distribution 
adequate for molecular beam experiments (solid line) \protect\cite{Gro95PRL}.
H$_2$ molecular beam adsorption experiment under normal incidence
(Rendulic {\it et al.}~\protect\onlinecite{Ren89}): circles;
H$_2$ effusive beam scattering experiment with an incident angle of
of $\theta_i = 15^{\circ}$
(Rettner and Auerbach~\protect\onlinecite{Ret96}): long-dashed line. }
\label{h2pdstick}
\end{figure}

Transition metal surfaces are usually very reactive as fas as
hydrogen dissociation is concerned \cite{Ren89,Aln89,Ber92,But94}.
In Fig.~\ref{h2pdstick} the results of molecular beam experiments
of Rendulic, Anger and Winkler \cite{Ren89} and of Rettner and
Auerbach \cite{Ret96} for the dissociative adsorption of H$_2$ 
on Pd(100) are shown. At low kinetic energies
these experiments yield a sticking probability above 0.5. But even 
more interestingly, the sticking probability initially decreases
with increasing kinetic energy. This is reminiscent of the dependence
of the sticking probability on the kinetic energy in atomic or
molecular adsorption illustrated in Fig.~\ref{mol_ads}. Therefore
for a long time it was believed that such an initially decreasing
sticking probability in dissociative adsorption is a signature of
the so-called precursor mechanism \cite{Ren94}. In this mechanism the 
molecule does not directly dissociate but is first trapped molecularly
in a precursor state from which it then dissociates. This trapping
probability decreases with increasing kinetic energy and thus
determines the sticking at low kinetic energies.

\begin{figure}[tb]
\unitlength1cm
\begin{center}
   \begin{picture}(10,11.0)
\put(-1.5,-0.5){ {\epsfysize=12.cm  
          \epsffile{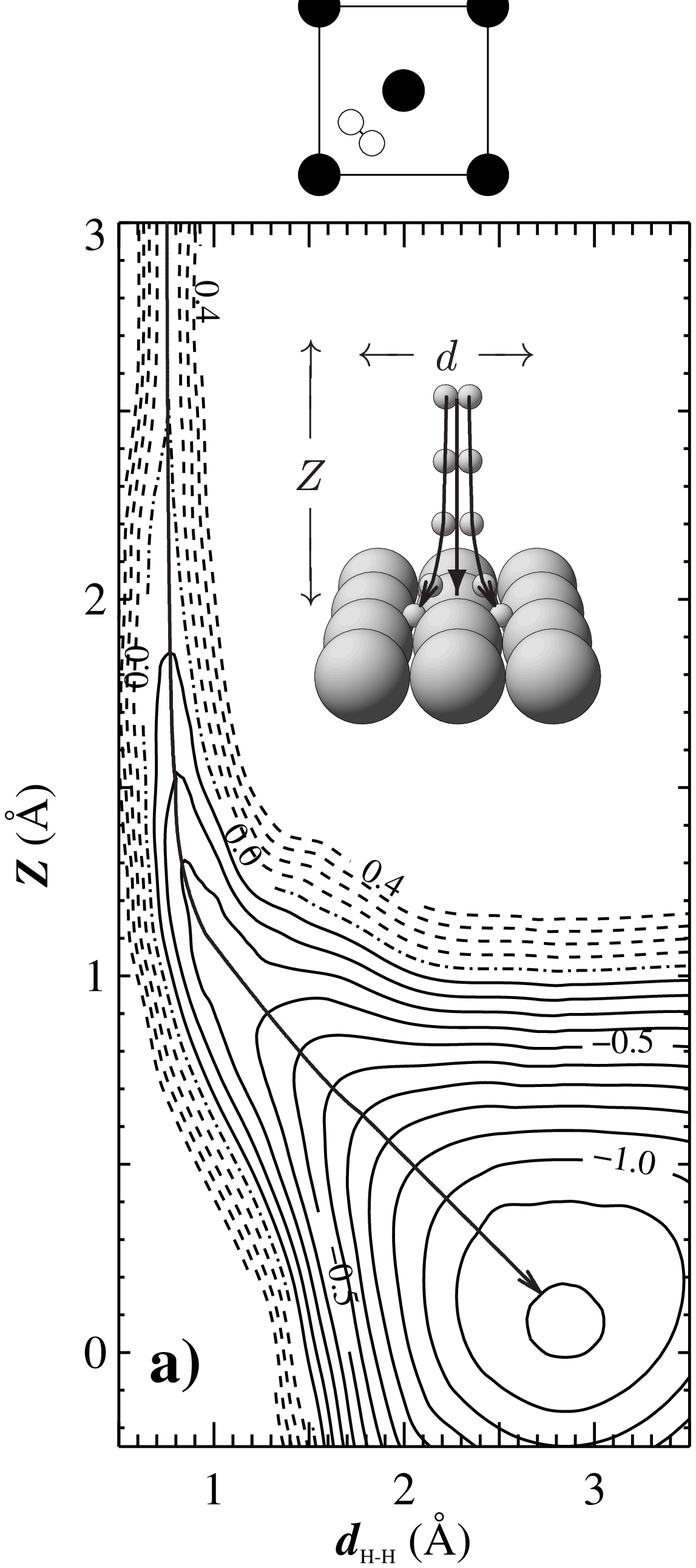}} }
\put(2.5,-0.5){ {\epsfysize=12.cm  
          \epsffile{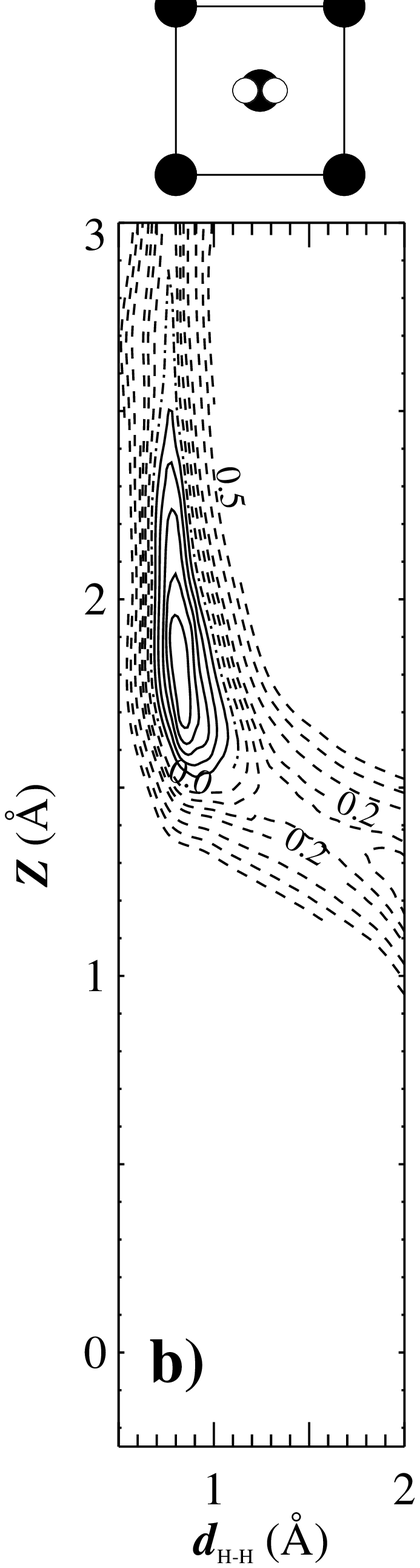}} }

   \end{picture}

\end{center}
   \caption{Contour plots of the PES along a two two-dimensional cuts 
            through the six-dimensional coordinate space of 
            H$_2$/Pd\,(100), so-called elbow plots, determined by GGA 
            calculations \protect\cite{Wil95,Wil96PRB}. The coordinates 
            in the figure are the H$_2$ center-of-mass distance 
            from the surface $Z$ and the H-H interatomic distance $d$.
            The dissociation process in the $Zd$ plane is illustrated
            in the inset. The lateral H$_2$ center-of-mass coordinates 
            in the surface unit cell and the orientation of the molecular 
            axis, i.e. the coordinates $X$, $Y$, $\theta$, and $\phi$  
            are kept fixed for each 2D cut and depicted above the 
            elbow plots. Energies are in eV per H$_2$ molecule.
            The contour spacing in a) is 0.1~eV, while it is 0.05~eV in b).}

\label{h2pdelbow}
\end{figure}

Wilke and Scheffler have performed density-functional theory (DFT) calculations
of the interaction of H$_2$ with Pd(100) in order to elucidate the dissociation
process. In Fig.~\ref{h2pdelbow} two so-called elbow plots are shown.
They represent a two-dimensional cut through the potential energy surface (PES)
of H$_2$/Pd(100) in which the orientation of the molecule, its 
center-of-mass lateral coordinates and the substrate are kept fixed. 
The molecule is oriented parallel to the surface, and the PES is plotted
as a function of the center-of-mass distance of the molecule from the
surface $Z$ and the interatomic distance $d_{\rm H-H}$. Fig.~\ref{h2pdelbow}a 
demonstrates that the dissociation of H$_2$ on Pd(100) is non-activated, i.e.,
there are reaction paths towards dissociative adsorption with no energy
barrier. The majority of reaction pathways, however, is hindered by
barriers \cite{Gro96CPLa}.   
Furthermore, in these calculations no molecular adsorption
state has been found. It looks like there is such a well in 
Fig.~\ref{h2pdelbow}b. However, the detailed DFT study of the PES
has shown that this apparent well does not correspond to a local minimum
of the PES, it is rather a saddle point in the multi-dimensional PES.

An analytical representation of this {\it ab initio} PES has been used
for a quantum dynamical study in which all six hydrogen degrees of freedom 
were taken into account explicitly while the substrate was kept fixed
\cite{Gro96CPLa}. The results are also plotted in Fig.~\ref{h2pdstick}.
The sticking curve for a monoenergetic beam initially in the vibrational and
rotational ground state shows a strong oscillatory structure which will be 
discussed below. An experimental molecular beam, however, does not correspond
to a monoenergetic beam in one specific quantum state. 
If one assumes an energy spread and a distribution of internal
molecular states typical for a beam experiment, the oscillations
are almost entirely smoothed out in the 6D quantum results
(solid line in fig.~\ref{h2pdstick}). The results corresponding to the
beam simulation agree with the experimental results semi-quantitatively.
More importantly, they reproduce the general trend found in the 
experiment, namely the initial decrease of the sticking probability
as a function of the kinetic energy followed by an increase at higher
energies. Now in the {\it ab initio} PES there is no molecular adsorption
state, furthermore in the 6D quantum dynamical calculation no energy
transfer to the substrate is considered. Hence the precursor mechanism
cannot be operative in the simulation. So what is the reason for the
initial decrease of the sticking probability?

Since energy transfer cannot play a crucial role in the adsorption process,
this decrease in the sticking probability has to be caused by a purely 
dynamical effect, namely the steering
effect \cite{Gro95PRL,Gro95JCP,King78,Kay95,Whi96}:
Although the majority of pathways to dissociative adsorption
has non-vanishing barriers with a rather broad distribution of
heights and positions, slow molecules can be very efficiently
steered to  non-activated pathways towards dissociative adsorption 
by the attractive forces of the potential. This mechanism becomes 
less effective at higher kinetic energies where the molecules are 
too fast to be focused into favourable configurations towards 
dissociative adsorption.  If the
kinetic energy is further increased, the molecules will eventually
have enough energy to directly traverse the barrier region leading
to the final rise in the sticking probability.

\begin{figure}[tb]
\unitlength1cm
\begin{center}
   \begin{picture}(10,6.5)
      \includegraphics{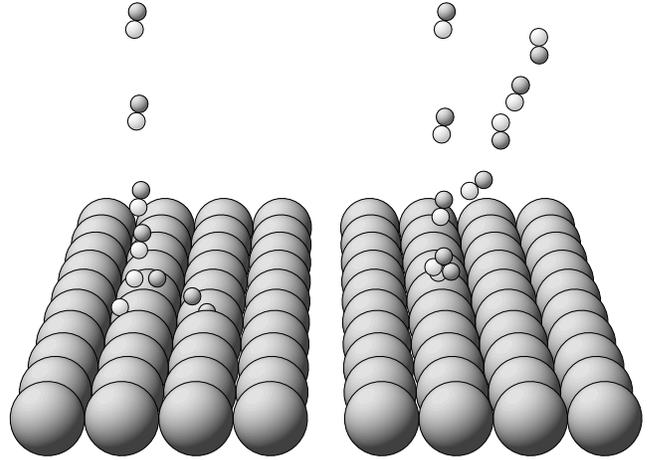}

   \end{picture}

\end{center}
   \caption{Snapshots of classical trajectories of hydrogen molecules 
impinging on a  Pd(100) surface. The initial conditions are chosen in 
such a way that the trajectories are restricted to the $xz$-plane.
Left trajectory: initial kinetic energy $E_i = 0.01$~eV. 
Right trajectory: same initial conditions as in the left trajectory
except that the molecule has a higher kinetic energy of 0.12 eV. }

\label{traj2run}
\end{figure}

In order to illustrate the steering effects, we use the results
of classical molecular dynamics calculations which have been
performed on exactly the same PES as the quantum dynamical calculations
\cite{Gro97Vac}. In these classical calculations significant
differences in the sticking probability compared to the quantum
results have been found which are mainly due to zero-point effects.
The steering effect, however, is a general mechanism operative in
quantum as well as in classical dynamics. I will therefore use
snapshots of two typical trajectories in order to illustrate
the dynamical mechanism responsible for the initial decrease
of the sticking probaiblity. These trajectories are plotted
in Fig.~\ref{traj2run}. The initial conditions are chosen in 
such a way that the trajectories are restricted to the $xz$-plane.
The left trajectory demonstrates why the sticking probability
is so large at low kinetic energies due to the steering effect.
The incident kinetic energy is $E_i = 0.01$~eV. In this particular 
example the molecular axis is initially almost perpendicular to the
surface. In such a configuration the molecule cannot dissociate
at the surface. But the molecule is so slow that the forces 
acting upon it can reorient the molecule. It is turned parallel 
to the surface and then follows a non-activated path towards 
dissociative adsorption. This shows how molecule with unfavorable
initial conditions can still dissociate due to very efficient
steering towards favorable configurations.

This process becomes less effective at higher kinetic energies,
which is demonstrated with the right trajectory in Fig.~\ref{traj2run}.
The initial conditions are the same as for the left trajectory, except  
for the higher kinetic energy of 0.12~eV. Of course the same forces
act upon the molecule, and due to the anisotropy of the PES the molecule also
starts to rotate to a configuration parallel to the surface. However,
now the molecule is too fast to finish this rotation. It hits the 
repulsive wall of the PES at the surface with its molecular axis tilted by 
about 45$^{\circ}$ with respect to the surface normal. At the classical 
turning point there is a very rapid rotation corresponding
to a flip-flop motion, and then the molecule is scattered back into 
the gas-phase rotationally excited.

\begin{figure}[tb]
\unitlength1cm
\begin{center}
   \begin{picture}(10,6.1)
      \includegraphics{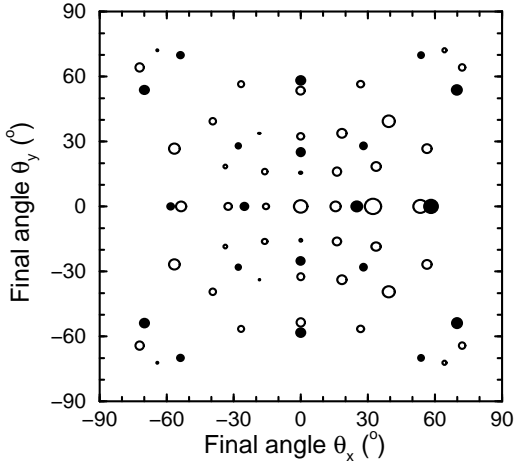}
   \end{picture}

\end{center}
   \caption{Angular distribution of the in-plane and out-of-plane
           of scattering of H$_2$/Pd(100). 
           The initial kinetic energy is $E_i = 76$~meV, the
           incident angle is $\theta_i = 32^{\circ}$ along the
           $\langle0 \bar 1 1\rangle$ direction. The molecules are initially
           in the rotational ground state $j_i = 0$.
           Open circles correspond to rotationally elastic, filled
           circles to rotationally inelastic diffraction.
           The radii of the circles are proportional to the logarithm 
           of the scattering intensity.  
           $x$ denotes the $\langle0 \bar 1 1\rangle$ direction, $y$ the
           $\langle0 1 1\rangle$ direction. The specular peak is the largest
           open circle (from Ref.~\protect\cite{Gro96CPLb}). }

\label{inout}
\end{figure} 

Now I like to come back to the strong oscillatory structure of the
sticking curve for a monoenergetic beam initially in one particular
quantum state in Fig.~\ref{h2pdstick}. 
These oscillations are a consequence of the quantum
nature of the hydrogen particle, in classical calculations they
do not appear \cite{Gro98PRB,Gro97Vac}. If a quantum particle is
interacting with a periodic surface, coherent scattering leads to
diffraction. Such a calculated diffraction pattern is shown in 
Fig.~\ref{inout} for hydrogen molecules in the ground state
scattering at a Pd(100) surface with an kinetic energy of $E_i = 76$~meV under 
an angle of incidence of $\theta_i = 32^{\circ}$. There are rather
many diffraction peaks since also rotationally inelastic diffraction 
can occur, i.e., scattering in which the rotational state of the
molecule is changed. Still the number of diffraction
peaks is finite and increases discontinously with increasing energy.
An analysis of the oscillatory structure of the sticking
probability in Fig.~\ref{h2pdstick} reveals that these oscillations
can be related to threshold effects associated with the opening of 
new scattering channels \cite{Gro96CPLb,Gro96PRL}.
These oscillations have not been observed experimentally yet,
although they have been carefully searched for \cite{Ret96,Ret96PRL}.
They are very sensitive to the symmetry of the initial conditions
\cite{Gro96CPLb,Gro96PRL}. Furthermore, since Pd(100) is a very
reactive surface, a large fraction of the incoming hydrogen molecules
is not scattered back coherently but adsorbs dissociatively. These adatoms
then disturb the periodicity of the surface and thus suppress,
in addition to already existing surface imperfections like steps
and vacancies, the coherence of the scattering events. Hence 
the experimental observation of this oscillations actually represents
a challenging task.

The dependence of adsorption and desorption 
on kinetic energy, molecular rotation and orientation
\cite{Gro95PRL,Gro96SSb}, molecular vibration \cite{Gro96CPLa}, ro-vibrational
coupling \cite{Gro96Prog}, angle of incidence \cite{Gro98PRB}, 
and the rotationally elastic and inelastic 
diffraction of H$_2$/Pd(100) \cite{Gro96CPLb} have been studied so far
by six-dimensional {\it ab initio} dynamics calculations
on the same PES. The results of these calculations have been compared 
to a number of independent experiments \cite{Ren89,Sch92,Beu95,Wet96}, and 
they are at least in semi-quantitative agreement with all of these experiments
This shows that {\it ab initio} dynamics calculations are indeed
capable of adequately describing the hydrogen dissociation
on transition metal surfaces.

\section{Dissociative adsorption at a sulfur-covered transition metal surface}

The presence of an adsorbate on a surface can profoundly change the 
surface reactivity. A well-known example is the reduction of the activity
of the car-exhaust catalyst by lead. But also adsorbed sulfur ``poisons''
this catalyst. An understanding of the underlying mechanisms and their 
consequences on the reaction rates is therefore of decisive importance
for, e.g., designing better catalysts. 
Traditionally an ``trial and error'' approach
was used to improve the activity of a catalyst by adding some
substances. 
On Pd(100) it is experimentally well-known that the presence of sulfur 
leads to a large reducting of hydrogen dissociation 
probability \cite{Ren89,Bur90}. While at the clean surface the dissociation
probability is about 60\% for a kinetic energy of $E_{\rm i} = 0.05$~eV,
at the sulfur-covered surface it drops below 1\% at the same
energy \cite{Ren89}.

Theoretically this problem had only been addressed by a small 
number of studies. These focused either on the 
adsorbate induced change of the density of states (DOS) at the Fermi level
\cite{Fei84,Fei85,MacL86} or on the adlayer induced electrostatic
field \cite{Nor84,Nor93,Ham93}. Just recently, the poisoning of hydrogen 
dissociation on Pd(100) by sulfur adsorption has been the subject of
detailed DFT studies \cite{Wil95,Wil96S,Wei97}.

\begin{figure*}[t]
\unitlength1cm
\begin{center}
   \begin{picture}(18,8.5)
\put(-1.,-4.){ \rotate[r]{\epsfysize=20.cm  
          \epsffile{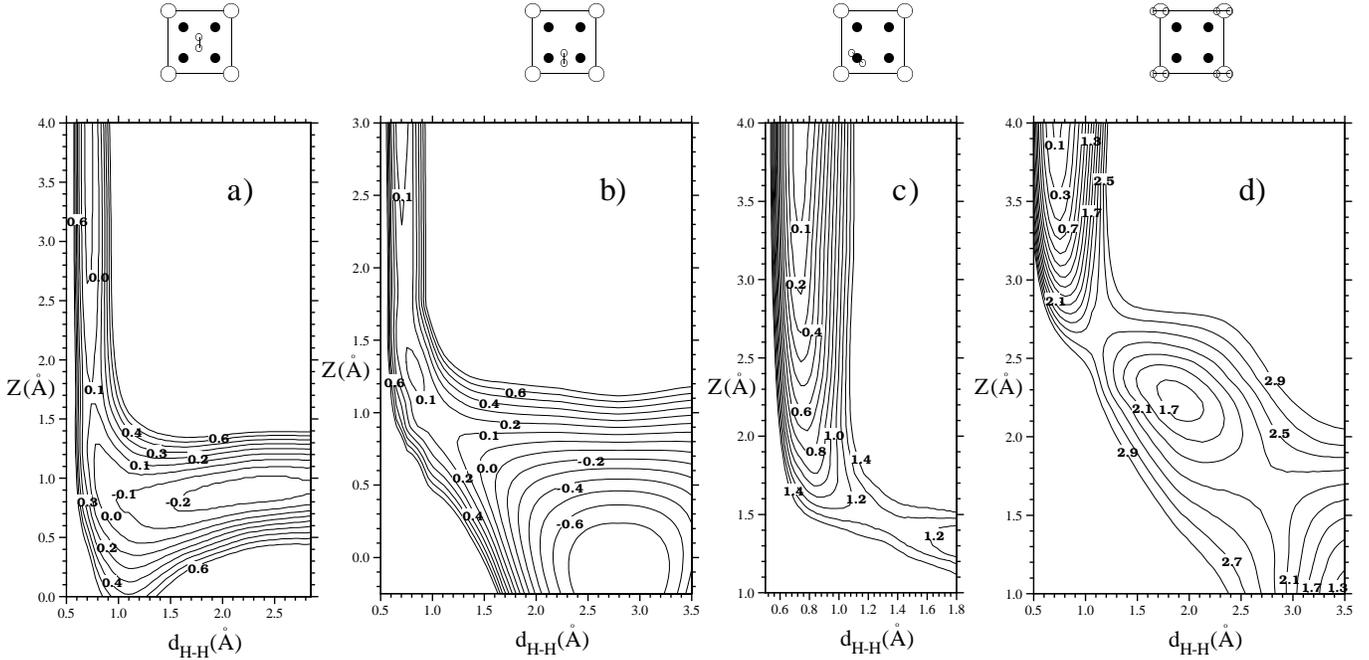}} }
   \end{picture}

\end{center}
\caption{ Cuts through the six-dimensional potential 
          energy surface (PES) of H$_2$ dissociation over
         (2$\times$2)S/Pd(100) at four different sites with
         the molecular axis parallel to the surface:
                a) at the fourfold hollow site;
                b) at the bridge site between two Pd atoms;
                c) on top of a Pd atom;
                d) on top of a S atom.
       The energy contours, given in eV per molecule,  
       are displayed as a function of the H-H distance, 
       $d_{H-H}$, and the height $Z$
       of the center-of-mass of H$_2$ above the topmost Pd layer. 
       The geometry of each dissociation pathway is indicated 
       in the panel above the contour plots. The large open circles
       are the sulfur atoms, the large filled circles are the
       palladium atoms.}

\label{H2SPd_PES}
\end{figure*}

In Fig.~\ref{H2SPd_PES} I have collected four elbow plots of the hydrogen
dissociation on the (2$\times$2) sulfur covered Pd(100) surface \cite{Wei97}.
In contrast to the clean Pd(100) surface, the hydrogen dissociation 
on the sulfur covered surface is no longer non-activated. The minimum
barrier, which is shown in Fig.~\ref{H2SPd_PES}a, has a height of
0.1~eV and corresponds to a configuration in which the H$_2$ center of
mass is located above the fourfold hollow site. This is the
site which is farthest away from the sulfur atoms in the surface unit cell.
Recall that the most favorable reaction path on the {\em clean} Pd(100)
surface corresponds to the H$_2$ molecule dissociating at the bridge position
between two Pd atoms (see Fig.~\ref{h2pdelbow}a). For this approach
geometry the dissociation at the sulfur covered surface is now
hindered by a barrier of height 0.15~eV (Fig.~\ref{H2SPd_PES}b). 
There is a peculiar local minimum at the dissociation path for this
configuration when the molecule is still 1\mbox{\AA} above the Pd atoms.
There are apparently subtle compensating effects between the attractive
interaction of H$_2$ with the Pd atoms and the repulsion originating
from the S atoms. Figs.~\ref{H2SPd_PES}a and b show 
that the dissociation is hindered by the formation of energy barriers
in the entrance channel of the potential energy surface, however, 
the hydrogen dissociation is still exothermic, i.e., the poisoning
is not due to site-blocking. This result is actually at variance
with measurements of the hydrogen saturation coverage as a function
of the sulfur coverage \cite{Bur90}.
In these experiments a linear decrease of the hydrogen saturation 
coverage with increasing sulfur coverages was found. At a sulfur coverage
of $\Theta_{\rm S} = 0.28$, which is close to the one used in the
calculations, hydrogen adsorption should be completely suppressed, i.e.,
there should be no attractive sites for hydrogen adsorption any more.
A possible explanation for this apparent contradiction will be given
below.

Closer to the sulfur atoms the PES becomes strongly repulsive.
This is illustrated in Fig.~\ref{H2SPd_PES}c and d. While the dissociation 
path over the Pd on-top position on the clean surface is hindered by a 
barrier of height 0.15~eV \cite{Wil96PRB} (Fig.~\ref{h2pdelbow}b), 
the adsorbed sulfur leads to an increase in this barrier height 
to 1.3~eV (Fig.~\ref{H2SPd_PES}c). Directly above the sulfur atoms 
the barrier towards dissociation even increases to values 
of about 2.5~eV for molecules oriented parallel to
the surface (Fig.~\ref{H2SPd_PES}d).

\begin{figure*}[t]
\unitlength1cm
\begin{center}
   \begin{picture}(18,6.5)
\put(-3.,-8.){ \rotate[r]{\epsfysize=24.cm  
          \epsffile{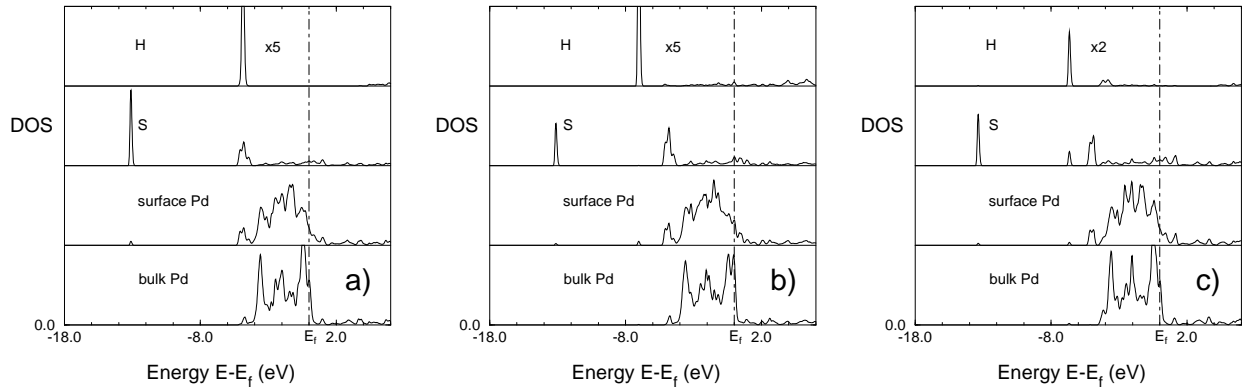}} }
   \end{picture}

\end{center}
   \caption{ Density of states (DOS) for a H$_2$ molecule situated at 
          (a) ($ Z,d_{\rm H-H}$)= (4.03\AA, 0.75\AA) and
          (b) ($Z,d_{\rm H-H}$)= (1.61\AA, 0.75\AA) above the fourfold
              hollow site which corresponds to the configuration
              depicted in Fig.~\protect\ref{H2SPd_PES}a,
              and for a H$_2$ molecule situated at
          (c) (Z,d$_{H-H}$)= (3.38\AA, 0.75\AA) above the sulfur
              atom which corresponds to the configuration
              depicted in Fig.~\protect\ref{H2SPd_PES}d.
              $Z$ and $d_{\rm H-H}$ denote the H$_2$ center-of-mass distance 
             from the surface and the H-H interatomic distance, respectively.
       Given is the local DOS at the H atoms, the S adatoms, 
       the surface Pd atoms, and the bulk Pd atoms. 
       The energies are given in eV.}

\label{DOS}
\end{figure*}

The goal of any theoretical study should be to provide a qualitative
picture that explains the calculated results. There are many different
ways of illustrating the electronic factors that determine the
reactivity of a particular system 
(see, e.g., Refs.~\cite{Wil96PRB,Eich96,Fei84,Ham95PRL,Wil96AP}). 
Current studies have emphasized that the reactivity of surfaces
cannot be solely understood by the electronic density of states
at the Fermi level \cite{Ham95,WCoh96}. In order to understand the origins of 
the formation of the small energy barriers at the hollow and bridge site and 
the large energy barriers at the top sites, we will therefore compare the 
whole relevant DOS for the H$_2$ molecule in these different geometries. 
For a discussion of the reactivity of the clean Pd(100) surface I refer
to Ref.~\cite{Wil96PRB}. Here I focus on the changes of the density of states 
induced by the presence of sulfur on the surface (Fig.~\ref{DOS}).

The information provided by the density of states alone is often
not sufficient to assess the reactivity of a particular system.
It is also important to know the character of the occupied and
unoccupied states. For the dissociation the occupation
of the bonding $\sigma_{\rm g}$ and the anti-bonding $\sigma_{\rm u}^*$ 
H$_2$ molecular levels {\em and} of the bonding and anti-bonding
states with respect to the surface-molecule interaction are of
particular importance.

Figure~\ref{DOS}a shows the DOS when the H$_2$ molecule is 
still far away from the surface above the fourfold hollow site,
i.e, in the configuration that corresponds to Fig.~\ref{H2SPd_PES}a.
The H-H distance $d$ is 0.75 \AA\  and the center of mass of the H$_2$ 
molecule is 4.03 \AA\ above the topmost Pd layer so that 
there is no interaction between the hydrogen molecule and 
the sulfur covered palladium surface.
The large peak in the sulfur DOS at -13~eV corresponds to the S 3$s$ state.
The sulfur {\it p} orbitals strongly interact with the Pd {\it d} states, 
which is evident from the peak in the sulfur DOS at the 
Pd {\it d} band edge (at E-E$_F$ = -4.8 eV) and from
the broad band at higher energies which has substantial 
weight close to the Fermi level. 
The {\it d} band at the surface Pd atoms is broadened and shifted
down somewhat with respect to the clean surface due to the interaction 
with the S atoms \cite{Wil96S}. There is one intense peak in the
hydrogen DOS at -4.8 eV which corresponds to the $\sigma_{\rm g}$ state.
This peak is degenerate with the sulfur related bonding state at -4.8 eV, 
this degeneracy, however, is accidental, as will become evident 
immediately.

The density of states for the molecule at the minimum barrier position
of Fig.~\ref{H2SPd_PES}a is shown in Fig.~\ref{DOS}b. Now the
$\sigma_{\rm g}$ state has shifted down to -7.1 eV while the
sulfur state at -4.8 eV remains almost unchanged. This indicates that
there is no direct interaction between hydrogen and sulfur. It also
proofs that the degeneracy between these two states in Fig.~\ref{DOS}a
is accidental. Furthermore, we find a broad distribution of hydrogen 
states with a small, but still significant weight below the Fermi level.
These are states of mainly H$_2$-surface antibonding character
\cite{Wil96S,Wei97}. A comparison with the hydrogen dissociation
at the clean Pd surface yields that more H$_2$-surface antibonding
states are populated at the sulfur covered surface. This is caused
by the sulfur induced downshift of the Pd $d$-band. These 
H$_2$-surface antibonding states lead to a repulsive interaction and thus 
to the building up of the barriers in the entrance channel of
the PES \cite{Wil96S}. It is therefore an indirect interaction
between sulfur and hydrogen that is responsible for the barriers
at this site. A similar picture explains why for example noble
metals are so unreactive for hydrogen dissociation: The low-lying
$d$-bands of the noble metals cause a downshift and a substantial 
occupation of the antibonding H$_2$-surface states resulting
in high barriers for hydrogen dissociation \cite{Ham95PRL}.

The situation is entirely different if the molecule approaches
the surface above the sulfur atom. This is demonstrated
in Fig.~\ref{DOS}c. The center of mass of the H$_2$ molecule is
still 3.38~\AA\ above the topmost Pd layer, but already at this
distance the hydrogen and the sulfur states strongly couple. The intense 
peak of the DOS at \mbox{-4.8 eV} has split into a sharp bonding state 
at -6.6 eV and a narrow anti-bonding state at -4.0 eV.
Thus it is a direct interaction of the hydrogen with the sulfur related
states that causes the high barriers towards hydrogen dissociation
close to the sulfur atoms. In conclusion, the poisoning of hydrogen
dissociation on Pd(100) by adsorbed sulfur is due to a combination
of a indirect effect, namely the sulfur-related downshift of the Pd $d$-bands 
resulting in a larger occupation of H$_2$-surface antibonding states,
with a direct repulsive interaction between H$_2$ and S close to the
sulfur atoms.  

\begin{figure}[tb]
\unitlength1cm
\begin{center}
   \begin{picture}(10,6.5)
\centerline{   \rotate[r]{\epsfysize=8.cm  
          \epsffile{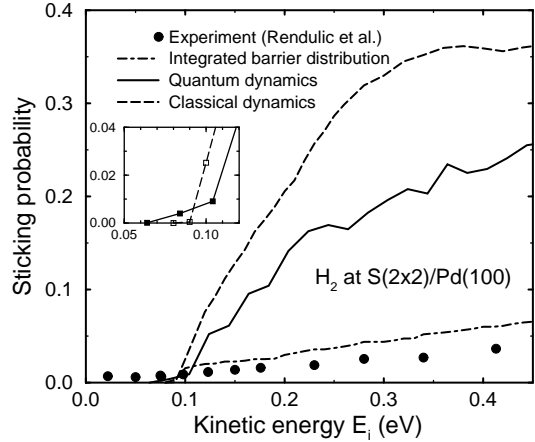}} }
   \end{picture}

\end{center}
   \caption{Sticking probability versus kinetic energy for
a H$_2$ beam under normal incidence on a S(2$\times$2)/Pd(100) surface.
Full dots: experiment (from ref.~\protect\cite{Ren89});
Dashed-dotted line: Integrated barrier distribution,
which corresponds to the sticking probability in the hole 
model \protect\cite{Kar87};
Solid line: Quantum mechanical results for molecules initially in the
rotational and vibrational ground-state;
Dashed line: Classical results for initially non-rotating and non-vibrating
molecules. The inset shows the quantum and classical results at low 
energies.}

\label{stick}
\end{figure}

In order to assess the dynamical consequences of the sulfur adsorption
on the hydrogen dissociation, six-dimensional dynamical calculations 
on the analytical representation of the {\it ab initio} PES of H$_2$ at 
S(2$\times$2)/Pd(100) have been performed \cite{Gro98PRL}.
The results of these quantum and classical calculations for the
H$_2$ dissociative adsorption probability as a function of the incident
energy are compared with the experiment \cite{Ren89} in Fig.~\ref{stick}.
In addition, also the integrated barrier distribution $P_b(E)$,
\begin{eqnarray} \label{barr}
P_b (E) & = & \frac{1}{2\pi A} \ \int 
\Theta (E - E_{\rm b} (\theta, \phi, X, Y)) \nonumber \\
& & \times \ \cos \theta d\theta \ d\phi \ dX \ dY
\end{eqnarray}
is plotted. Here $\theta$ and $\phi$ are the polar and azimuthal
orientation of the molecule, $X$ and $Y$ are the lateral coordinates of 
the hydrogen center-of-mass. $A$ is the area of the surface unit cell.
Each quadruple defines a cut through the six-dimensional space (see
Fig.~\ref{H2SPd_PES} for examples), 
and $E_{\rm b}$ is the minimum energy barrier 
along such a cut. The function $\Theta$ is the Heavyside step function. 
The quantity $P_b(E)$ is the fraction 
of the configuration space for which the barrier towards dissociation
is less than $E$. If there were no steering effects, $P_b(E)$ would
give the classical sticking probability, i.e., it corresponds to the 
sticking probability in the classical sudden approximation or the so-called 
``hole model'' \cite{Kar87}.

First of all it is evident that the calculated sticking probabilities are 
significantly larger than the experimental results. 
Only the onset of dissociative adsorption at $E_{\rm i} \approx 0.12$~eV 
is reproduced by the calculations. This onset is indeed also in agreement
with the experimentally measured mean kinetic energy of hydrogen 
molecules desorbing from sulfur covered Pd(100) \cite{Com80}.
The question arises where these large differences between theory and
experiment come from. It might be that uncertainties in the experimental 
determination are responsible for the difference. 
The exact sulfur coverage in the 
experiment was not very well characterized. The sulfur adlayer was obtained 
by simply heating up the sample which leads to segregation of bulk sulfur 
at the surface. The sulfur coverage was then monitored
through the ratio of the Auger peaks S$_{132}$/Pd$_{330}$ \cite{Ren89}.
Since the hydrogen sticking probability depends sensitively on the
sulfur coverage \cite{Ren89,Bur90}, a small uncertainty in the
sulfur coverage can have a decisive influence. However, as noted above,
while the DFT calculations yield that the poisoning is caused by
the building up of barriers hindering the dissociation, the vanishing
hydrogen saturation coverage for roughly a quarter monolayer of adsorbed
sulfur \cite{Bur90} suggests that any attractive adsorption sites 
for hydrogen have disappeared due to the presence of sulfur.
These seemingly contradicting
results and also the discrepancy between calculated and measured molecular
beam sticking probabilities could be reconciled if subsurface sulfur 
plays an important role for the hydrogen adsorption energies.
Subsurface sulfur is not considered in the calculations but 
might well be present in the experimental samples. The possible
influence of subsurface species on reactions at surfaces certainly
represents a very interesting and important research subject
for future investigations.

Except for this open question, there are further interesting
results obtained by the dynamical calculations. The calculated
sticking probabilities are not only much larger than the
experimental ones, they are also much larger than what one would
expect from the hole model. This demonstrates that steering is
not only operative for potential energy surfaces with non-activated
reaction paths like for H$_2$/Pd(100), but also for activated
systems as H$_2$/S(2$\times$2)/Pd(100). As Fig.~\ref{H2SPd_PES}   
demonstrates, the sulfur covered Pd surface represents a 
strongly corrugated system with barrier heights varying by
more than 2~eV for molecules with their axis parallel to the
surface. And the large barriers above the sulfur atoms extend
rather far into the gas phase (see Fig.~\ref{H2SPd_PES}d).
Molecules with unfavorable initial conditions are very effectively
reoriented to low-barrier sites \cite{Gro98PRL}. This leads to
an enhancement of the sticking probability with respect to the
hole model by a factor of three to four.

Figure \ref{stick} shows furthermore that the classical molecular
dynamics calculations over-estimate the sticking probability 
of H$_2$ at S(2$\times$2)/Pd(100) compared to the quantum results. 
At small energies below the minimum barrier height the quantum calculations 
still show some dissociation due to tunneling, as the inset of 
Fig.~\ref{stick} reveals, whereas the classical results are of course zero. 
But for higher energies the classical sticking probability is up to 
almost 50\% larger than the quantum sticking probabilities. 
This suppression is also caused by the large corrugation and the
anisotropy of the PES. The wave function describing the 
molecule has to pass narrow valleys in the PES in the angular and lateral 
degrees of freedom in order to dissociate. This leads to a localization
of the wave function and thereby to the building up of zero-point energies
which act as additional effective barriers. While
the vibrational H-H mode becomes softer upon dissociation so that
the zero-point energy in this particular mode decreases,
for the system H$_2$/S(2$\times$2)/Pd(100) this
decrease is over-compensated by the increase in the zero-point
energies of the four other modes perpendicular to the reaction
path, i.e., the sum of {\em all} zero-point energies increases
upon adsorption \cite{Gro98PRL}. Therefore the quantum particles
experience an effectively higher barrier region causing the
suppressed sticking probability compared to the classical
particles. Interestingly enough, if the sum of all zero-point energies
remains approximately constant along the reaction path as in the
system H$_2$/Pd(100), then these quantum effects almost cancel 
out \cite{Gro98PRB,Gro97Vac}.

\section{Conclusions}

In this review {\it ab initio} studies of reactions
on surfaces have been presented. In the last years the interaction between
electronic structure calculations on the one side and dynamical calculations
on the other side has been very fruitful. The availability of high-dimensional
reliable potential energy surfaces has challenged the dynamics community
to improve their methods in order to perform high-dimensional dynamical 
studies on these potentials. Now quantum studies of the dissociation of 
hydrogen on surfaces are possible in which all six degrees of 
freedom of the molecule are treated dynamically. In this review I have
tried to show that this achievement represents an important step forward 
in our understanding of the interaction of molecules with surfaces.
Not only the quantitative agreement with experiment is improved, but
also important qualitative concepts emerge from the electronic structure
calculations as well as from the high-dimensional dynamical simulations.
These concepts are applicable to any reaction system. This represents the 
importance of hydrogen as a model system for studying reactions on surfaces.
However, the {\it ab initio} treatment of reactions on surfaces has now 
matured enough so that in the future there will be also more studies
on other reaction systems like for example the important class of
oxidation reactions on surfaces. This will be the next step towards
a full microscopic description of catalytic reactions, one of the
ultimate goals of surface science.

\section*{Acknowledgements}

I am very grateful to my collegues and coworkers who have made this work 
possible. I would like to mention in particular Thomas Brunner, 
Bj{\o}rk~Hammer, Ralf~Russ, Ching-Ming Wei, Steffen~Wilke, 
and last but not least my supervisors
Helmar Teichler, Wilhelm Brenig and Matthias Scheffler.


\begin{thebibliography}{99}

\bibitem{Ham94} B. Hammer, M. Scheffler, K.W. Jacobsen, and J.K. N{\o}rskov,
                Phys. Rev. Lett. {\bf 73}, 1400 (1994).
\bibitem{Whi94} J.A. White, D.M. Bird, M.C. Payne, and I. Stich,
                Phys. Rev. Lett. {\bf 73}, 1404 (1994).

\bibitem{Wil95} S. Wilke and M. Scheffler, Surf. Sci. {\bf 329}, L605 (1995).
\bibitem{Wil96PRB} S. Wilke and M. Scheffler, 
                  Phys. Rev. B {\bf 53}, 4926 (1996).

\bibitem{Whi96PRB} J.A. White, D.M. Bird, and M.C. Payne, Phys. Rev. B 
                  {\bf 53}, 1667 (1996).

\bibitem{Wie96} G. Wiesenekker, G.J. Kroes, and E.J. Baerends, J. Chem. Phys.
               {\bf 104}, 7344 (1996).


\bibitem{Eich96} A. Eichler, G. Kresse, and J. Hafner, Phys. Rev. Lett. 
                 {\bf 77}, 1119 (1996).

\bibitem{Dong96} W. Dong and J. Hafner, Phys. Rev. B {\bf 56}, 15396 (1996).

\bibitem{Gro95PRL} A. Gross, S. Wilke, and M. Scheffler, Phys. Rev. Lett.
                  {\bf 75}, 2718 (1995).
\bibitem{Gro98PRB} A. Gross and M. Scheffler, Phys. Rev. B {\bf 57}, 2493
                   (1998).

\bibitem{Kro97PRL} G.J. Kroes, E.J. Baerends, and R.C. Mowrey,
                 Phys. Rev. Lett. {\bf 78}, 3583 (1997).
\bibitem{Kro97JCP} G.J. Kroes, E.J. Baerends, and R.C. Mowrey,
                  J. Chem. Phys. {\bf 107}, 3309 (1997).
\bibitem{Dai97} J. Dai and J.C. Light, J. Chem. Phys. {\bf 107},
                1676 (1997).


\bibitem{Eilm96} G. Eilmsteiner, W. Walkner, and A. Winkler,
                 Surf. Sci. {\bf 352}, 263 (1996).
\bibitem{Schu83} G. Schulze and M. Henzler, Surf. Sci. {\bf 124}, 336 (1983).
 
\bibitem{Ren89} K.D. Rendulic, G. Anger, and A. Winkler, Surf. Sci. {\bf 208},
                404 (1989).

\bibitem{Bra96PRB} P. Bratu, W. Brenig, A. Gross, M. Hartmann, U. H{\"o}fer
        P. Kratzer, and R. Russ, Phys. Rev. B {\bf 54}, 5978 (1996).

\bibitem{Hin97}    O. Hinrichsen, F. Rosowski, A. Hornung, M. Muhler,
                  and G. Ertl, J. Catal. {\bf 165}, 33 (1997). 

\bibitem{Zan88} A. Zangwill, Physics at Surfaces (Cambridge University Press,
               Cambridge, 1988). 

\bibitem{Sch88} H. Schlichting, D. Menzel, T. Brunner, W. Brenig, 
                and J.C. Tully, Phys. Rev. Lett. {\bf 60}, 2515 (1988).

\bibitem{Gro94PRL} A. Gross, B. Hammer, M. Scheffler, and W. Brenig, Phys. Rev.
Lett. {\bf 73}, 3121 (1994).

\bibitem{Ret95} C.T. Rettner, H.A. Michelsen, and D.J. Auerbach, J. Chem. Phys.
                {\bf 102}, 4625 (1995).


\bibitem{Aln89} P. Alnot, A. Cassuto, and D.A. King, Surf. Sci. {\bf 215},  
                29 (1989).
\bibitem{Ber92} H.F. Berger, Ch. Resch, E. Gr\"osslinger, G. Eilmsteiner, A.
Winkler, and K.D. Rendulic, Surf. Sci. {\bf 275}, L627 (1992)
\bibitem{But94} D.A. Butler, B.E. Hayden, and J.D. Jones, Chem. Phys. Lett.
{\bf 217}, 423 (1994).

\bibitem{Ret96} C.T. Rettner and D.J. Auerbach, Chem. Phys. Lett. {\bf 253},
236 (1996).

\bibitem{Ren94} K.D. Rendulic and A. Winkler, Surf. Sci. {\bf 299/300}, 261
(1994). 


\bibitem{Gro96CPLa} A. Gross and M. Scheffler, Chem. Phys. Lett. {\bf 256},
417 (1996).

\bibitem{Gro95JCP} A. Gross, J. Chem. Phys. {\bf 102}, 5045 (1995).


\bibitem{King78} D.A. King, CRC Crit Rev. Solid State Mater. Sci. {\bf 7}, 167
            (1978).

\bibitem{Kay95} M. Kay, G.R. Darling, S. Holloway, J.A. White, and D.M. Bird,
              Chem. Phys. Lett. {\bf 245}, 311 (1995).
\bibitem{Whi96} J.A. White, D.M. Bird, and M.C. Payne, Phys. Rev. B {\bf 53},
             1667 (1996).


\bibitem{Gro97Vac} A. Gross and M. Scheffler, J. Vac. Sci. Technol. A15,
1624 (1997). 



\bibitem{Gro96CPLb} A. Gross and M. Scheffler, Chem. Phys. Lett.
{\bf 263}, 567 (1996). 

\bibitem{Gro96PRL} A. Gross and M. Scheffler, Phys. Rev. Lett. {\bf 77},
405 (1996).


\bibitem{Ret96PRL} C.T. Rettner and D.J. Auerbach,  Phys. Rev. Lett. {\bf 77},
404 (1996).


\bibitem{Gro96SSb} A. Gross, S. Wilke, and M. Scheffler, Surf. Sci. 
{\bf 357/358}, 614 (1996). 

\bibitem{Gro96Prog} A. Gross and M. Scheffler, Prog. Surf. Sci. {\bf 53}, 187
(1996). 



\bibitem{Sch92} L. Schr{\"o}ter, S. K{\"u}chenhoff, R. David, W.~Brenig, and
                H.~Zacharias, Surf. Sci. {\bf 261}, 243 (1992). 


\bibitem{Beu95} M. Beutl, M. Riedler, and K.D. Rendulic, Chem. Phys. Lett.
                {\bf 247}, 249 (1995).

\bibitem{Wet96} D. Wetzig, R. Dopheide, M. Rutkowski, R. David, and
H. Zacharias, Phys. Rev. Lett. {\bf 76}, 463 (1996).



\bibitem{Bur90} M.L. Burke and R.J. Madix, Surf. Sci. {\bf 237}, 1 (1990).


\bibitem {Fei84} P.J. Feibelman and D.R. Hamann, Phys. Rev. Lett. {\bf 52}, 
                 61 (1984).
\bibitem {Fei85} P.J. Feibelman and D.R. Hamann, Surf. Sci. {\bf 149}, 
                48 (1985).
\bibitem {MacL86} J.M. MacLaren, J.B. Pendry, and R.W. Joyner, 
                  Surf. Sci. {\bf 165}, L80 (1986).


\bibitem{Nor84} J.K. N{\o}rskov, S. Holloway, and N.D. Lang, Surf. Sci. 
               {\bf 137}, 65 (1984).
\bibitem {Nor93} J.K. N{\o}rskov, in 
{\it The Chemical Physics of Solid Surfaces}, edited by 
D.A. King and D.P. Woodruff (Elsevier, Amsterdam, 1993), Vol. 6, p. 1.
\bibitem {Ham93} B. Hammer, K.W. Jacobsen, and J.K. N{\o}rskov, 
                 Surf. Sci. {\bf 297}, L68 (1993).


\bibitem{Wil96S} S. Wilke and M. Scheffler, Phys. Rev. Lett. {\bf 76}, 
                 3380 (1996).
\bibitem{Wei97} C.M. Wei, A. Gross, and M. Scheffler, Phys. Rev. B, submitted.



\bibitem{Ham95PRL} B. Hammer and M. Scheffler, Phys. Rev. Lett. {\bf 74},
                   3487 (1995). 

\bibitem{Wil96AP} S. Wilke, Appl. Phys. A {\bf 63}, 583 (1996).
\bibitem{Ham95} B. Hammer and J.K. N{\o}rskov, Surf. Sci. {\bf 343}, 
                211 (1995).
\bibitem {WCoh96} S. Wilke, M. Cohen, and M. Scheffler, Phys. Rev. Lett. 
                  {\bf 77}, 1560 (1996).


\bibitem{Gro98PRL} A. Gross, C.M. Wei, and M. Scheffler, submitted.

\bibitem{Kar87} M. Karikorpi, S. Holloway, N. Henriksen and J.K. N{\o}rskov, 
                Surf. Sci. {\bf 179}, L41 (1987).



\bibitem{Com80} G. Comsa, R. David, and B.-J. Schumacher, 
                Surf. Sci. {\bf 95}, L210 (1980).





\end{thebibliography}
\end{document}